\documentclass[prl,aps,twocolumn,superscriptaddress,preprintnumbers,asmath,amssymb]{revtex4}
\usepackage{color}
\usepackage{graphicx}
\usepackage{ulem}
\usepackage{float}
\usepackage{here}
\usepackage{hyperref}

\UseRawInputEncoding

\begin{document}
\title{In-plane Isotropy of the Low Energy Phonon Anomalies in YBa$_{2}$Cu$_{3}$O$_{6+x}$}

\author{Sofia-Michaela Souliou}
\affiliation{Institute for Quantum Materials and Technologies, Karlsruhe Institute of Technology, 76131 Karlsruhe, Germany \\}
\author{Kaushik Sen}
\affiliation{Institute for Quantum Materials and Technologies, Karlsruhe Institute of Technology, 76131 Karlsruhe, Germany \\}
\author{Rolf Heid}
\affiliation{Institute for Quantum Materials and Technologies, Karlsruhe Institute of Technology, 76131 Karlsruhe, Germany \\}

\author{Suguru Nakata}
\affiliation{Max Planck Institute for Solid State Research, Heisenbergstr. 1, D-70569 Stuttgart, Germany \\}

\author{Lichen Wang}
\affiliation{Max Planck Institute for Solid State Research, Heisenbergstr. 1, D-70569 Stuttgart, Germany \\}

\author{Hun-ho Kim}
\affiliation{Max Planck Institute for Solid State Research, Heisenbergstr. 1, D-70569 Stuttgart, Germany \\}

\author{Hiroshi Uchiyama}
\affiliation{Japan Synchrotron Radiation Research Institute (JASRI/SPring-8), 1-1-1 Kouto, Sayo, Hyogo 679-5198 Japan}

\author{Michael Merz}
\affiliation{Institute for Quantum Materials and Technologies, Karlsruhe Institute of Technology, 76131 Karlsruhe, Germany \\}\author{Matteo Minola}
\affiliation{Max Planck Institute for Solid State Research, Heisenbergstr. 1, D-70569 Stuttgart, Germany \\}

\author{Bernhard Keimer}
\affiliation{Max Planck Institute for Solid State Research, Heisenbergstr. 1, D-70569 Stuttgart, Germany \\}

\author{Matthieu Le Tacon}
\affiliation{Institute for Quantum Materials and Technologies, Karlsruhe Institute of Technology, 76131 Karlsruhe, Germany \\}


\begin{abstract}We study the temperature dependence of the low energy phonons in the $(H, 0, L)$ reciprocal plane of the highly ordered ortho-II YBa$_2$Cu$_3$O$_{6.55}$ cuprate high temperature superconductor by means of high-resolution inelastic x-ray scattering. Anomalies associated with the emergence of long-range charge density wave (CDW) fluctuations are observed, and are qualitatively similar to those previously observed in the $(0, K, L)$ plane. This confirms the unconventional nature of this bi-dimensional CDW, which is not soft-phonon driven. With the support of first principles calculations, the symmetry of the anomalous phonon is identified and is found to match that of the charge modulation. This suggests in turn that these anomalies originate from a direct coupling between the phonons and the collective CDW excitations.
\end{abstract}

\maketitle

\section{Introduction}
The role of the electron-phonon interaction (EPI) in the singular physical properties of high temperature superconducting cuprates has been debated since their discovery. 
Although its driving influence on the superconducting instability has been rapidly discarded, its impact on other electronic phases remains largely unclear~\cite{Frano2020,Keimer2015}.
The discovery of ubiquitous charge density waves (CDW) in the cuprates a decade ago, has shed a new light on this issue, although the associated phonon anomalies~\cite{Pintschovius2002, Reznik2006, Raichle2011, Reznik2012, Blackburn2013,LeTacon2014, Miao2018} deviate substantially from those encountered in canonical €˜Peierls€™ CDW systems~\cite{Renker1974, Hoesch2009}.
In these materials, the structural distortion is driven by the condensation of soft-phonons at a reciprocal space wavevector where the electronic susceptibility diverges~\cite{Johannes2006}.
In the cuprates, phonon anomalies have been first reported on the high energy branches~\cite{Pintschovius2002, Reznik2006}, and associated with the formation of dynamical charge stripes. More recently, fingerprints of the coupling between the lattice and the charge degrees of freedom have been observed in the low energy lattice dynamics~\cite{Blackburn2013,LeTacon2014, Miao2018}.

More specifically, inelastic x-ray scattering (IXS) experiments on YBa$_{2}$Cu$_{3}$O$_{6+x}$ (YBCO$_{6+x}$) have shown strong renormalizations of the low energy phonons as quasi-two-dimensional (2D) short-range charge correlations appear in the CuO$_{2}$ planes at temperatures below $T_{CDW}$, which far exceeds the superconducting transition temperature $T_{c}$~\cite{LeTacon2014,Blackburn2013,Souliou2018}. 
The low energy phonon anomalies have been observed for different doping levels of the YBCO$_{6+x}$ family (x=0.55, 0.67 and 0.75) and are absent for the non-modulated YBCO$_{7}$. 
They therefore appear as a benchmark for the presence of 2D-CDW modulations. 
The observed anomalies include a pronounced broadening of a low energy acoustical phonon peak, appearing below $T_{CDW}$ and reaching its maximum value at $T_{c}$.
At lower temperatures, a sharp and deep softening of the phonon energy was observed in the superconducting state, reminiscent of the Kohn anomalies of soft-mode driven CDWs. 
Unlike in those systems, however, the anomaly weakens upon cooling further in the superconducting state. 
Both anomalies are sharply localized in momentum space around the CDW ordering wavevectors, testifying to a strong and very anisotropic EPI.

\begin{figure}
	\includegraphics[width=0.3\textwidth]{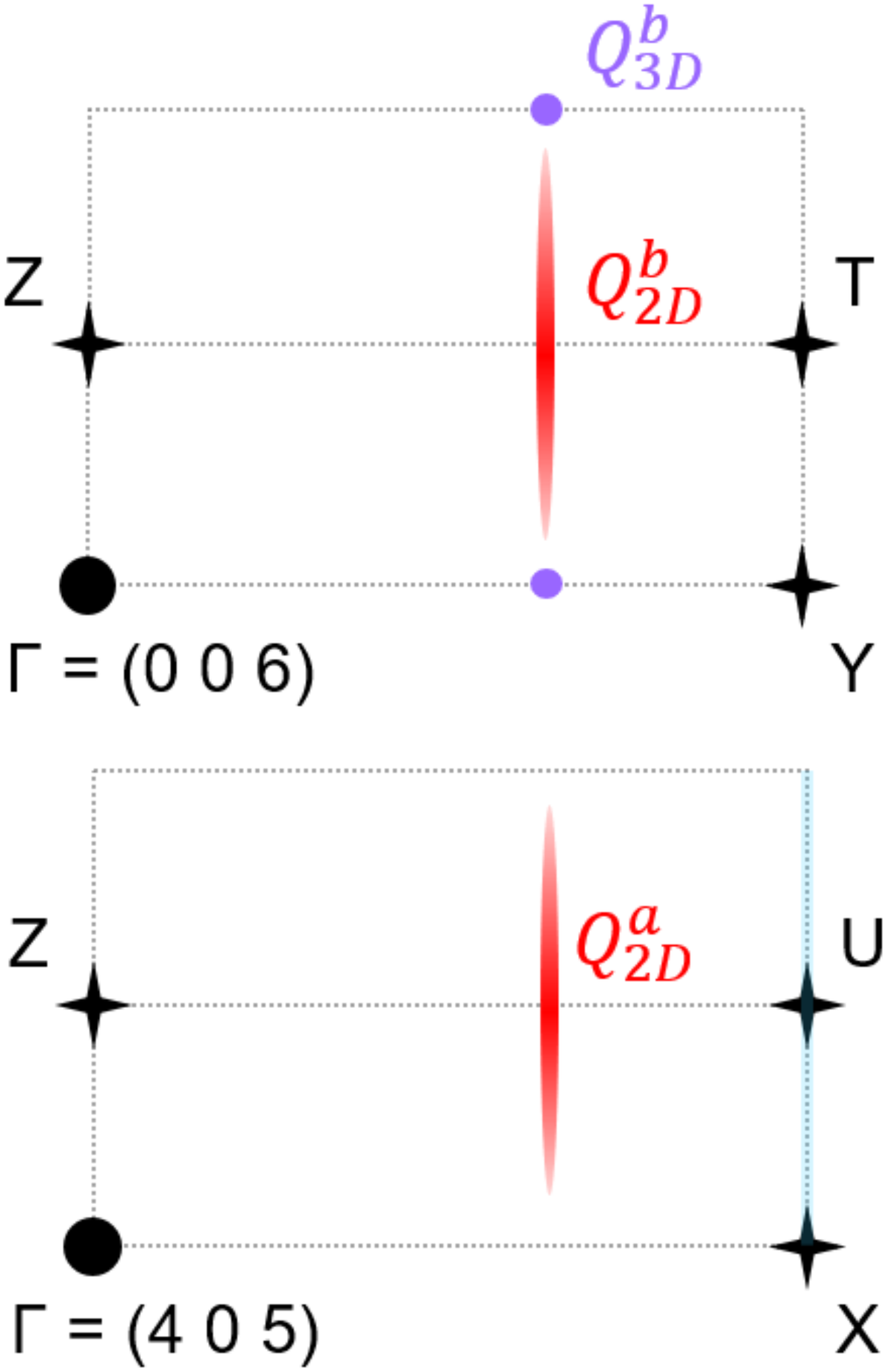}
	\centering
	\caption{(Color online) Schematic representation of the $(0, K, L)$ and $(H, 0, L)$ reciprocal lattice plane regions where the IXS measurements of references ~\cite{LeTacon2014, Kim2018} and the current measurements were performed respectively. The elongated satellite peaks corresponding to the $Q_{2D}^a$ and $Q_{2D}^b$ order are illustrated in red, whereas the satellite peaks corresponding to the $Q_{3D}^b$ order appearing under uniaxial compression or high magnetic field are illustrated in purple. The cut along which the superstructure peaks associated with the ortho-II oxygen ordering appear is highlighted in cyan. The corners of the dashed grey rectangle correspond to the high symmetry points of the Brillouin zones at which the measurements were performed.}
	\label{f1}
\end{figure}

The 2D-CDW in YBCO$_{6+x}$ is described by two modulation vectors, lying on the ($H$,0,$L$) and (0,$K$,$L$) planes respectively. It is weakly but clearly peaked at half-integer $L$ values (see Fig. \ref{f1}), indicative of a doubling of the unit cell in this direction and has very low correlation lengths along the $c$-axis, hence it is considered as a 2D-CDW~\cite{Chang2012}.
Both the incommensurability of the two ordering vectors and the relative intensities of the respective satellite CDW diffraction peaks exhibit an in-plane doping-dependent anisotropy.
In the case of YBCO$_{6.55}$ the two CDW ordering vectors are $q_{2D}^a$ =(0.315, 0 , 0.5) and $q_{2D}^b$=(0, 0.325, 0.5), with the satellites at $q_{2D}^a$ being weaker than the ones at $q_{2D}^b$ (as is generally the case).~\cite{Blackburn2013b,BlancoCanosa2014,Hucker2014}
However, the presence of an intense x-ray signal of structural origin (arising from the partially depleted Cu-O chains in the ($H$,0,$L$) reciprocal-space plane of YBCO$_{6+x}$) has limited the investigation of the low energy lattice dynamics to the (0,$K$,$L$) plane, i.e. only around $q_{2D}^b$~\cite{LeTacon2014,Blackburn2013}.
Recently, experiments under pressure~\cite{Souliou2018, Vinograd2019, Souliou2020} have been used to gain further insights regarding the interplay between the CDW and superconductivity. 
In particular, it has been shown that a large uniaxial compression along the $a$-axis, which suppresses superconductivity, induces a long-range 3D-CDW, with an in-plane incommensurability same as the one of the 2D-CDW, albeit with much longer correlation lengths along the c-axis (see the schematic representation in Fig. \ref{f1})~\cite{Kim2018,Kim2021}. Similar observations have been previously reported in the presence of a magnetic field sufficiently intense to suppress superconductivity~\cite{Gerber2015,Jang2016,Choi2020}.
Lattice dynamics studies revealed that the formation of this new phase, which has only been observed at $q_{3D}^b$=(0, 0.315, 1) both under field and stress - is associated with the condensation of an optical mode~\cite{Kim2018}. 
Interestingly, under uncompressed conditions the superconductivity-induced Kohn anomaly was detected along the entire (0, 0.315, $L$) reciprocal space cut connecting $q_{2D}^b$ and $q_{3D}^b$, raising novel questions regarding the interplay between the phonon anomalies, and the 2D and 3D charge orders in YBCO$_{6+x}$.

Both the in-plane anisotropy of the 2D-CDW and the unidirectional nature of the 3D-CDW suggest an intrinsic difference between the $a$- and $b$- axis modulations of the CuO$_2$ planes. 
On the other hand, the symmetric response of the two modulations to uniaxial stress~\cite{Kim2021} indicates a common underlying mechanism.
To provide a more solid foundation for the theoretical modeling of the CDW formation, it would be particularly useful to gain additional insights regarding the EPI around $q_{2D}^a$. 
Here we tackle this challenge directly, through a detailed high-resolution IXS study of the low energy lattice dynamics of YBCO$_{6+x}$ in the $(H, 0, L)$ plane.

Our experimental results reveal pronounced phonon anomalies at $q_{2D}^a$, strikingly similar to those observed at $q_{2D}^b$, both above and below the superconducting transition temperature. Our results complete the overview of the low energy lattice dynamics in charge ordered YBCO$_{6+x}$ and demonstrate a qualitatively comparable response in the $(H, 0, L)$ and $(0, K, L)$ reciprocal planes. The symmetry of the anomalous phonons can be determined by confronting our experimental results with those of density functional perturbation theory (DFPT) calculations, which reveal the same $B_1$ symmetry, as that inferred for the CDW from earlier diffraction experiments~\cite{Forgan2015}. This allows us to discuss a possible mechanism for the anomalous phonon behavior in YBCO$_{6+x}$. 

\begin{figure}
	\includegraphics[width=0.4\textwidth]{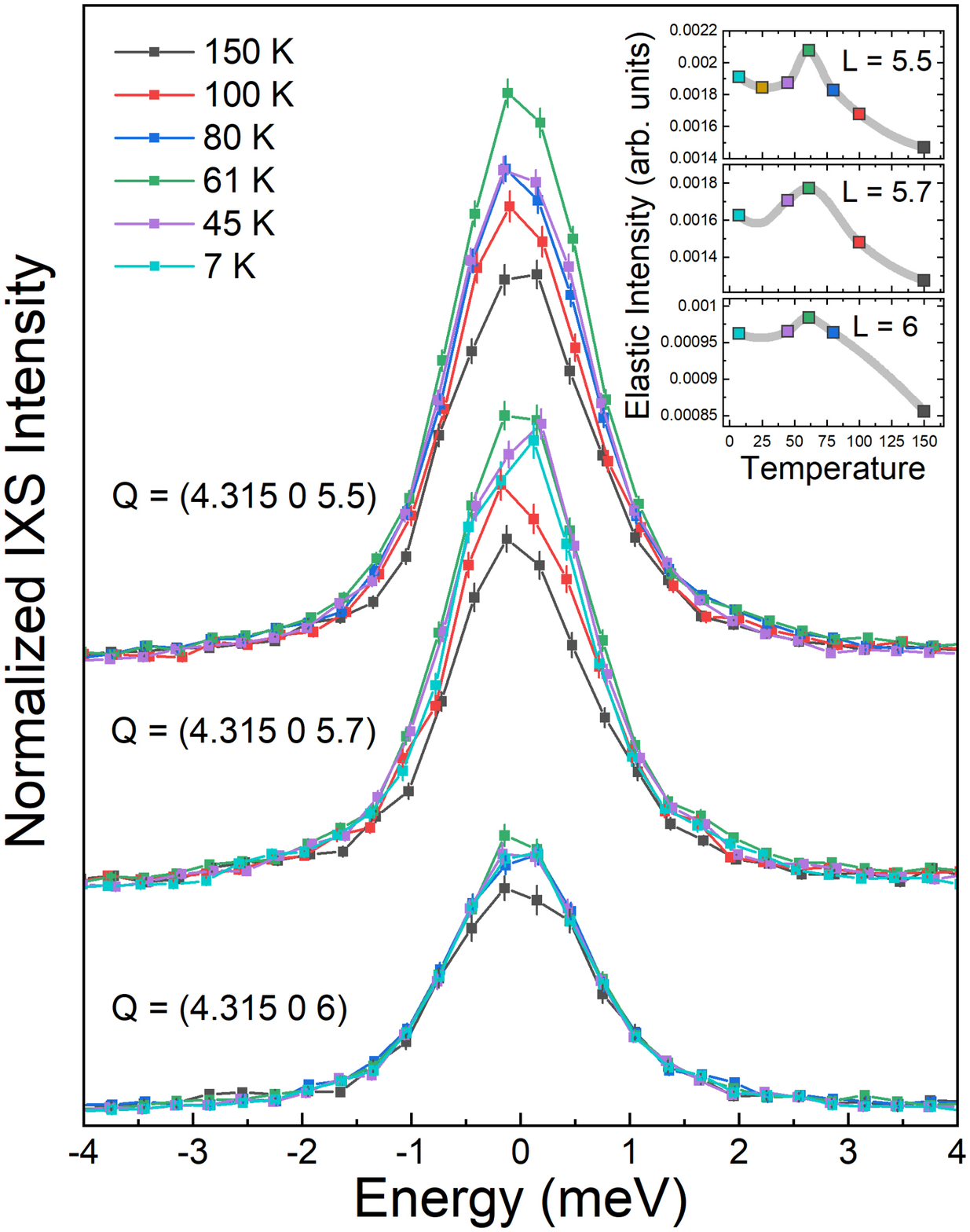}
	\centering
	\caption{(Color online) Temperature dependence of the IXS spectra around the central elastic peak at Q = $Q_{2D}^a$, Q= (4.315 0 5.7) and Q = (4.315 0 6). The spectra at different Q-positions are vertically shifted for clarity. The insets show the temperature dependence of the integrated intensity of the central peak at $L$ = 5.5, 5.7 and 6. The thick grey lines are guides to the eye.}
	\label{f2}
\end{figure}

\section{Experimental Methods}

For our study we used high-quality single crystals of YBa$_{2}$Cu$_{3}$O$_{6+x}$ grown by a flux method as described elsewhere. ~\cite{Lin2002} 
The oxygen content was adjusted to x=0.55 (\textit{p} = 0.114, $T_c$ = 61 K) through an appropriate annealing procedure and the samples were subsequently mechanically detwinned by heating under uniaxial stress. The samples were manually polished in a platelet-shape and etched with diluted HCl to minimize the surface damage contribution to the scattering. 
We have chosen to work with samples of this specific oxygen content in order to ensure a significant $q$-space separation between the 2D-CDW superstructure reflections at $q_{2D}^a$ = (0.315, 0, 0.5) and the much stronger reflections arising from the ordering of full and empty Cu-O chains sequences along the a-axis.~\cite{Andersen1999,Zimmermann2003,Strempfer2004}
YBa$_{2}$Cu$_{3}$O$_{6.55}$ has an ortho-II Cu-O chain ordering pattern with the relevant superlattice peaks appearing at q$_{ortho-II}$ = (0.5, 0, 0), conveniently away from $q_{2D}^a$.
The ortho-II oxygen ordering and the twin-free nature of the studied sample were confirmed by x-ray diffraction measurements (see Fig. \ref{App_xrd}).

The IXS experiments were performed at the BL35XU beamline of the SPring-8 (Japan).~\cite{Baron2000} The incident x-ray beam had an energy of 21.747 keV and a corresponding instrumental resolution of 1.5 meV. The beam was focused on an 80 $\times$ ~50~$\mu$m spot on the sample surface. Throughout this paper, the momentum transfers are quoted in reciprocal lattice units (r.l.u.) of the orthorhombic crystal structure (\textit{Pmmm} space group, $a$ = 3.8312(5) $\AA$, $b$ = 3.8731(4) $\AA$ , $c$ = 11.7331(10) $\AA$ at 295 K - note that the structural refinement was made with the twice larger value of the $a$ parameter in the ortho-II unit cell - see Appendix \ref{xrd}).

The sample was mounted on a closed-cycle cryostat with the $a^*$ and $c^*$ reciprocal space directions close to the horizontal scattering plane and the $b^*$ direction vertical and perpendicular to it. The momentum resolution was set to 0.016, 0.066 and 0.048 r.l.u. along the $a^*$, $b^*$ and $c^*$ directions, respectively. 

\section{Results and Discussion}

The IXS measurements were performed close to the G$_{405}$ = (4 0 5) Bragg reflection. 
The choice of this particular Brillouin zone was based on the results of x-ray diffraction experiments which reported a strong CDW sattelite peak at $Q_{2D}^a$= G$_{405}$ + $q_{2D}^a$ = (4.315 0 5.5) ~\cite{Forgan2015} and the results of DFPT caclulations according to which substantial structure factors are expected for the low energy phonon modes in the vicinity of $Q_{2D}^a$. 
We first report on the observation of the CDW satellite in this Brillouin zone by showing in Fig. \ref{f2} the IXS spectra recorded at $Q_{2D}^a$ around zero energy loss for selected temperatures. 
An enhancement of the elastic peak intensity is observed over a narrow $q$-range around $Q_{2D}^a$ along both the $H$ and the $K$ directions of the reciprocal space (see also the momentum dependence of the elastic intensity across $Q_{2D}^a$ plotted in Fig. \ref{App_1}) when cooling towards the superconducting $T_c$, and is followed by its partial suppression upon further cooling in the superconducting state. 
Along the $L$-direction, a clear enhancement of the elastic intensity is observed at $L$=5.7, while at $L$=6 the elastic peak increases only slightly upon cooling.  
We note that in all cases the enhanced elastic line seen at low temperatures is energy-resolution limited.
Away from $Q_{2D}^a$, the elastic peak intensity - which mostly arises from incoherent scattering due to the presence of lattice defects - is essentially temperature independent in all parts of the momentum space which we have explored in our study.
The behavior of the elastic line of the IXS spectra is therefore perfectly in line with previous observations in the perpendicular $(0, K, L)$ plane of YBa$_{2}$Cu$_{3}$O$_{6+x}$. 
In analogy with previous work on displacive phase transitions~\cite{Halperin1976}, this central peak can be associated to the pinning of fluctuating CDW nanodomains on defects ~\cite{LeTacon2014,Blackburn2013}.

\begin{figure}
	\includegraphics[width=0.48\textwidth]{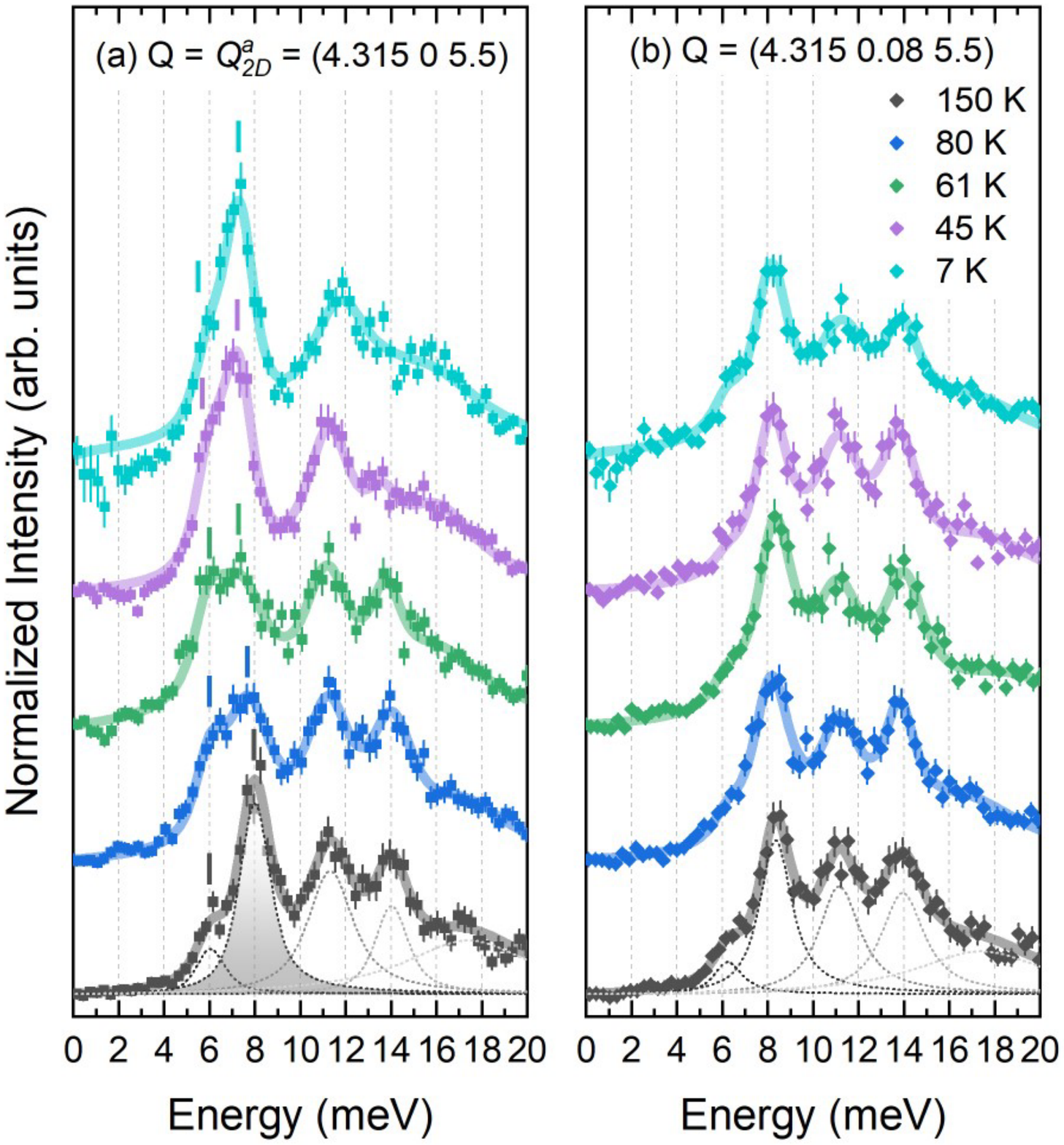}
	\caption{(Color online) (a,b) Inelastic part of the IXS spectra at (a) Q = $Q_{2D}^a$ and (b) Q$_{off-K}$ = (4.315 0.08 5.5) at selected temperatures. Thick solid lines correspond to the results of the least square fitting of the data,  the dotted lines indicate the contributions from the individual fitted phonon profiles and the ticks mark the energies of the two lowest energy phonon modes. The strongly renormalizing $B_1$ phonon mode is highlighted with the shadowed area in (a).}
	\label{f3}
\end{figure}

We now turn to the inelastic part of the IXS spectra, obtained after subtraction of the elastic line, and the Bose factor correction. We show these for $Q_{2D}^a$ and slightly away from it (at Q$_{off-K}$ = (4.315 0.08 5.5)) in Fig. \ref{f3}.
The IXS spectra at $Q_{2D}^a$ and Q$_{off-K}$ are rather similar at 150 K, as expected from their reciprocal space proximity.
Within the working energy resolution, we can clearly resolve four phonon lines below 15 meV. 
As shown in the Appendix, the number of modes as well as their frequencies are in good agreement with our first-principles structure factor calculations, which were performed considering the backfolding arising from the ortho-II superstructure (Fig. \ref{App_2}). 
The calculation indicates that the IXS intensity above $\sim$16 meV originates from multiple, weak phonon branches that will not be further discussed. 
As illustrated in the lower spectrum of each of the panels of Fig. \ref{f3}, we have analyzed the low energy part of the spectra by fitting the data with Lorentzian lineshapes, convoluted with the experimental resolution function.

In striking contrast to the spectra recorded at Q$_{off-K}$, which remain essentially unmodified upon cooling down to base temperature, drastic changes are observed in the low temperature IXS spectra at $Q_{2D}^a$.
In particular, next to an essentially temperature independent, weak feature near 6 meV associated with the O-superstructure backfolding (see Fig. \ref{App_2}), the strong $\sim$8 meV phonon mode (highlighted in Fig. \ref{f3}) is strongly renormalized both upon cooling from 150 K to $T_c$ and in the superconducting state.
The initially almost energy resolution limited phonon profile continuously broadens in the charge ordered state reaching a full-width at half-maximum (FWHM) of $\sim$1.5 meV at $T_c$, while its energy continuously softens to $\sim$7.3 meV (see discussion below). 
In the superconducting state the phonon profile abruptly changes, narrowing again to a resolution limited FWHM.
We reemphasize here that these low energy phonon anomalies are localized in the momentum space around $Q_{2D}^a$ both along the $K$ and along the $H$ direction of the momentum space (IXS spectra at Q = (4.25 0 5.5) and Q = (4.3 0 5.5) are given in Fig. \ref{App_3} and show the momentum sharpness of the phonon anomalies) and can thereby be unambiguously associated with the formation of the 2D-CDW.

\begin{figure}
	\includegraphics[width=0.42\textwidth]{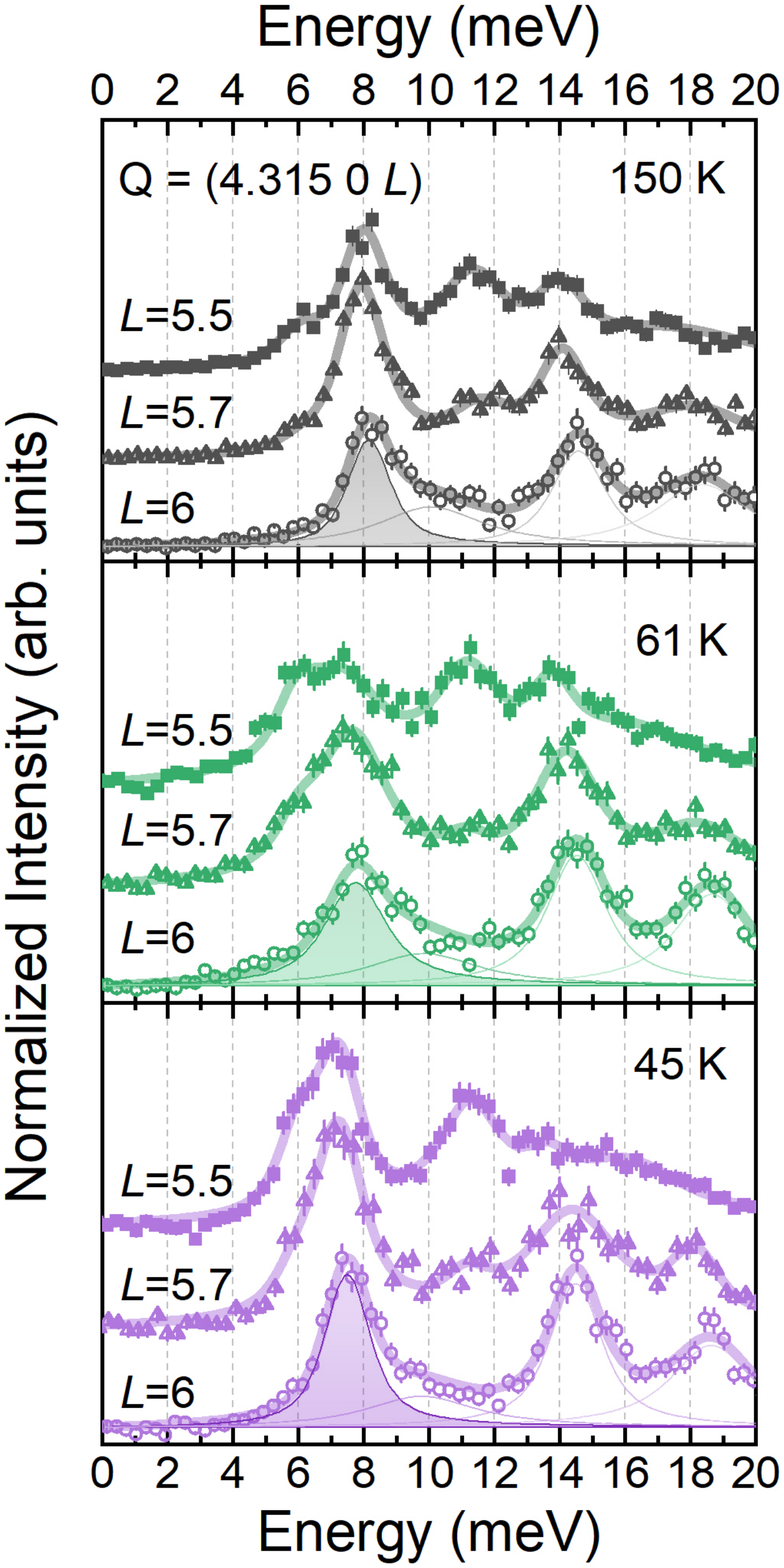}
	\centering
	\caption{(Color online) $L$ dependence of the IXS spectra at selected temperatures. The IXS spectra were recorded at Q = $Q_{2D}^a$, Q =(4.315 0 5.7) and Q =(4.315 0 6). Thick solid lines correspond to the results of the least square fitting of the data and the thin lines indicate the contributions from the individual fitted phonon profiles. The strongly renormalized $B_1$ phonon mode is highlighted.}
	\label{f4}
\end{figure}

\begin{figure}
	\includegraphics[width=0.45\textwidth]{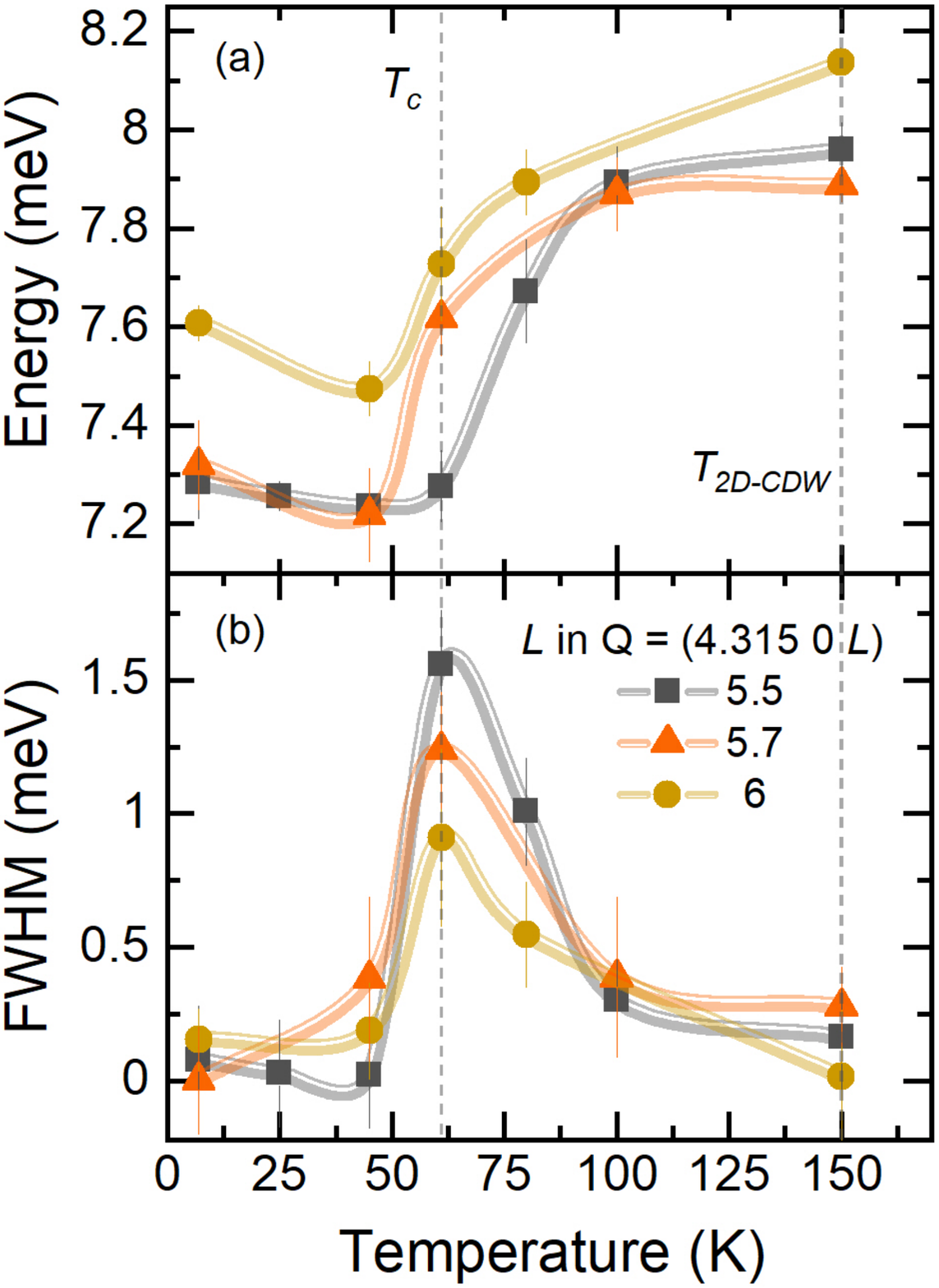}
	\centering
	\caption{(Color online) Temperature dependence of the energy and FWHM of the $B_{1}$ acoustical phonon (highlighted with the shadowed area in Fig. \ref{f3} and Fig. \ref{f4}) at Q = $Q_{2D}^a$, Q =(4.315 0 5.7) and Q =(4.315 0 6).}
	\label{f5}
\end{figure}

On the contrary, the anomalies are observed along the full span of the Brillouin zone in the perpendicular $L$ direction, as illustrated in Fig. \ref{f4}, where we show phonon spectra recorded at $Q_{2D}^a$, Q = (4.315 0 5.7) and Q = (4.315 0 6).
Pronounced renormalizations of the $\sim$8 meV phonon mode are observed in all of the above reciprocal space positions, both below the onset temperature of the 2D-CDW and below $T_c$.
The detailed temperature and $L$-dependence of the phonon energy and linewidth are given in Fig. \ref{f5}.
The observed phonon broadening continuously decreases in magnitude when moving from $Q_{2D}^a$ towards integer $L$. 
Within our resolution, the phonon softening remains almost unchanged in absolute value along the $L$ direction.

Our main finding is therefore that the phonon behavior in the two perpendicular $(H, 0, L)$ and $(0, K, L)$ planes is qualitatively very similar. 
In both cases it is very unusual with respect to the commonly encountered phononic response of conventional, soft-mode driven, CDW systems~\cite{Renker1974, Hoesch2009, Weber2011}, in which sharp Kohn anomalies are observed at the CDW ordering vector and the phase transition to the CDW state occurs when the soft-phonon frequency reaches zero. 
In YBCO$_{6+x}$, no anomalies in the low-energy phonon dispersion can be detected at (or above) the temperature $T_{CDW}$ at which the CDW satellites appears.  
As no thermodynamic phase transition can further be associated with this temperature, it has long be concluded that these satellites were in fact the fingerprints of long-range CDW fluctuations, whose condensation is prevented by the appearance of the competing superconducting phase. 
This view is supported by the formation of a true long-range CDW order as superconductivity is suppressed by means of high magnetic fields~\cite{Gerber2015,Jang2016,Choi2020} or uniaxial stress~\cite{Kim2018,Kim2021}. 
In the latter case the complete softening of an optical phonon has furthermore been directly observed~\cite{Kim2018}. Future investigation of the stress dependence of the lattice dynamics in the $(H,0,L)$ plane should be most insightful to understand the absence of a 3D-CDW in the $(H, 0, L)$ plane or to determine the actual experimental conditions under which it may form.

Compared to previous datasets, the higher energy resolution used for the present study enables us to draw a more accurate picture of the temperature dependence of the phonon anomaly. 
Simultaneously with the broadening of the $\sim$8 meV phonon near $Q_{2D}^a$ below $T_{CDW}$, we identify a softening of the low energy phonons. 
Both the broadening and the softening are maximized close to the superconducting $T_c$. 
Interestingly, the detailed temperature dependence of the phonon renormalizations exhibit a slight dependence along $L$.  

From a group symmetry analysis, it can be shown that the renormalized mode corresponds to an acoustic phonon branch belonging to the $B_{1}$ representation in agreement with previous findings in the $(0, K, L)$ plane of charge-ordered YBCO$_{6+x}$ ~\cite{LeTacon2014, Blackburn2013, Souliou2018, Kim2018}. 
The corresponding vibration pattern at the incommensurate wavevector of the CDW is very complex, but is in particular characterized by a symmetry of the atomic displacements against the mirror plane defined by the Cu-O chains. 
Most importantly, it should be noted that the symmetry operations associated with the anomalous $B_1$ mode (which are listed in the table~\ref{tab:symmetry}) are identical to that of the high energy bond-stretching mode $\Delta_1$ and, unlike that of modes in the $B_4$ representation, are fully compatible with ionic displacement associated with the CDW as inferred from the structural refinement on another ortho-II YBCO crystal~\cite{Forgan2015}. 
This enables us to propose a mechanism for the singular linewidth dependence of the phonon. 
It is clear that the appearance of the charge modulation is not driven by a phonon softening as in a conventional CDW. 
Irrespective of its microscopic origin, once established, the CDW and the associated collective excitations, such as those inferred from inelastic neutron~\cite{Park2014} or resonant inelastic x-ray~\cite{Chaix2017, Lee2021} scattering will hybridize with the phonons of compatible symmetry, similarly to the mechanism previously proposed by Kaneshita et al. in the context of charge stripes~\cite{Kaneshita2002}. 
As the phonon linewidth measured with IXS directly relates to the phonon self-energy, such coupling naturally yields a broadening
(i.e. a decrease of these phonons' lifetime) close to $Q_{2D}^a$. The opening of the superconducting gap in the particle-hole continuum below $T_c$, which is relatively large at this doping level ($\Delta \sim$ 30 meV~\cite{LeTacon2006}), suppresses this scattering channel and restores a resolution limited linewidth for the low-energy modes, leaving however unaffected those at higher energies, such as the bond-stretching modes above 70 meV which display sizeable anomalies down to the lowest temperatures~\cite{Pintschovius2002, Reznik2006, Reznik2012}.
%
%
\section{Summary and Conclusions}
In summary, we have studied the low energy lattice dynamics in the so-far unexplored $(H, 0, L)$ plane of YBa$_{2}$Cu$_{3}$O$_{6.55}$. 
Pronounced phonon renormalizations were observed around the CDW ordering vector across both the CDW and superconductivity onset temperatures. 
These phonon anomalies are qualitatively similar to those previously reported in the orthogonal $(0, K, L)$ plane, indicating the absence of a pronounced in-plane anisotropy in the low energy phonon response to the 2D-CDW order. 
The symmetry of the anomalous phonons could be identified and matches that of the charge modulation, enabling a direct coupling between them.
\newline
\section*{Acknowledgements}
This work was supported by the SPring-8 under the proposals numbers 2019A1203 and 2019B1345. 
Self-flux growth was performed by the Scientific Facility "Crystal Growth" at the  Max Planck Institute for Solid State Research, Stuttgart, Germany. The contribution from M.M. was supported by the Karlsruhe Nano Micro Facility for Information (KNMFi).

\appendix
\section{ Structural characterization of our single-crystalline YBa$_{2}$Cu$_{3}$O$_{6.55}$ sample}
\label{xrd}

\begin{figure}
	\includegraphics[width=0.45\textwidth]{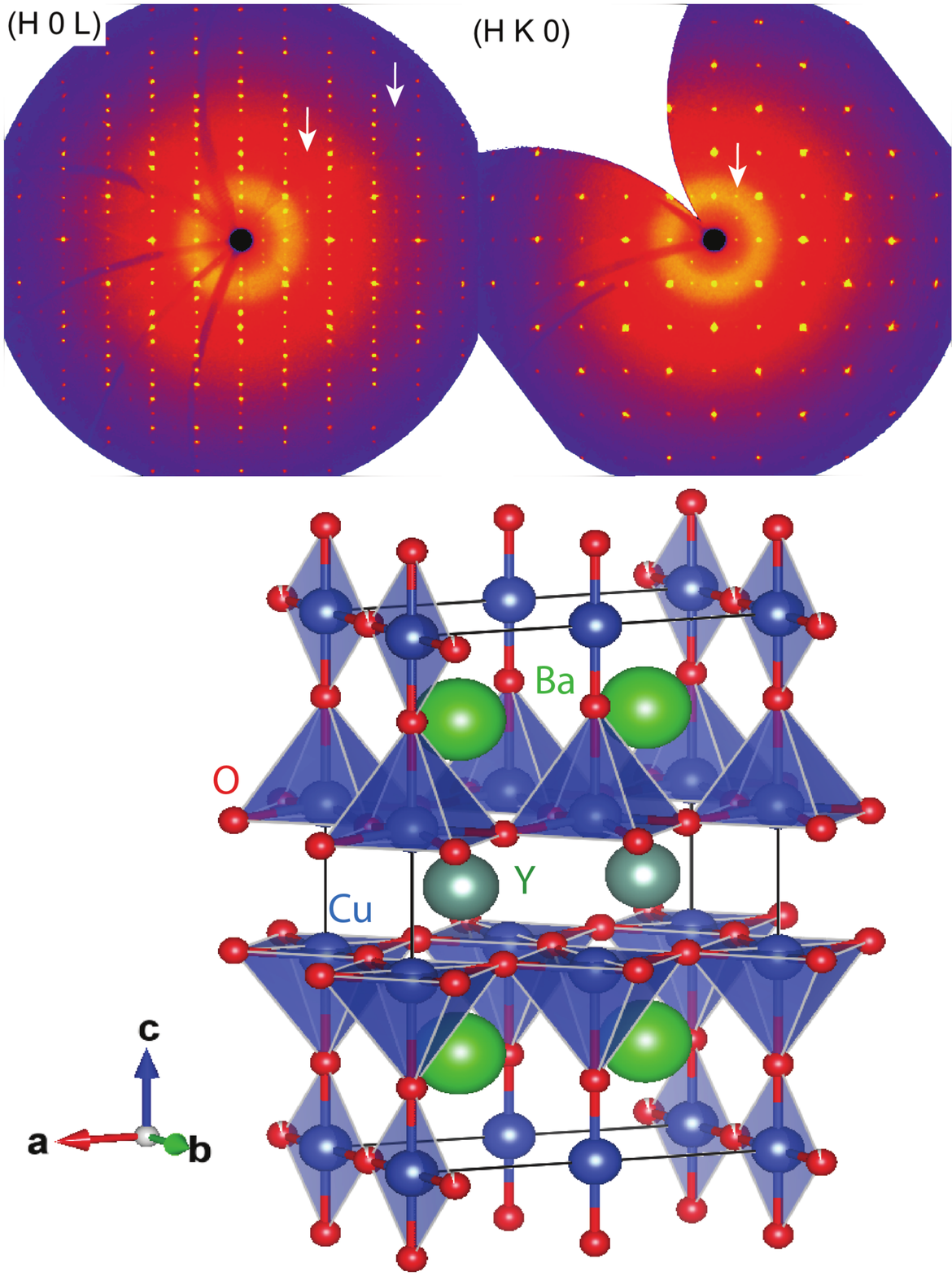}
	\centering
	\caption{(Color online) Maps of the $(H, 0, L)$  and ($H$, $K$, 0) reciprocal lattice planes of our YBa$_2$Cu$_3$O$_{6.55}$ single crystal. Selected ortho-II superstructure reflections at ($H$+0.5 0 $L$) are indicated by the white arrows. The unit cell of the ortho-II structure is depicted in the lower panel.}
	\label{App_xrd}
\end{figure}

X-ray diffraction (XRD) data on our YBa$_{2}$Cu$_{3}$O$_{6.55}$ single crystal sample were collected at 295 and 100 K on a STOE imaging plate diffraction system (IPDS-2T) using Mo $K_{\alpha}$ radiation. All accessible symmetry-equivalent reflections ($\approx 5950$) were measured up to a maximum angle of $2 \Theta =65°$.\@ The data were corrected for Lorentz, polarization, extinction, and absorption effects. Reconstructions of the reciprocal $(H, 0, L)$ and ($H$, $K$, 0) planes are depicted in Fig. \ref{App_xrd} and are clearly showing the superstructure peaks of the ortho-II phase measured at 295 K (see arrows). The unit cell of the ortho-II structure is also illustrated in Fig. \ref{App_xrd}.

Using SHELXL~\cite{Sheldrick2008} and JANA2006~\cite{Petricek2014}, around 695 averaged symmetry-independent reflections ($I > 2 \sigma$) have been included for structure determination and for the corresponding refinements in the orthorhombic space group (SG) $Pmmm$.\@ The structure was refined for the smaller averaged unit cell as well as for the doubled unit cell including the ortho-II chain oxygen ordering. In both cases the oxygen content of YBa$_{2}$Cu$_{3}$O$_{x}$ was determined to $x =6.47(3)$.\@ The refinements converged quite well and show very good reliability factors (see GOF, $R_1$,\@ and $wR_2$ in Table \ref{tab:table1}).

\begin{table}[h]
\begin{ruledtabular}
		\begin{tabular}[b]{rrcccc}
    &    &  &  295 K & 100 K & WP \\
			\hline
	&		&  $a_{ortho-II} = 2a$ (\r{A}) & 7.6625(8) & 7.6582(13) \\
	&		&  $b$ (\r{A}) & 3.8731(4) & 3.8693(6) \\
	&		&  $c$ (\r{A}) & 11.7331(10) & 11.6976(17)  \\[1mm]
	& Y		& $x$ & 0.25114(12) & 0.25125(13) & $2l$  \\
	&		& $U_{\rm eqiv}$ (\r{A}$^2$) & 0.00543(13) & 0.00221(16)   \\[1mm]
	& Ba	& $x$ & 0.24574(4) & 0.24551(4) & $4x$  \\
	&		& $z$ & 0.18864(2) & 0.18863(3) &   \\
	&		& $U_{\rm eqiv}$ (\r{A}$^2$) & 0.00734(6) & 0.00247(8) \\[1mm]
	& Cu1 & $U_{\rm eqiv}$ (\r{A}$^2$) & 0.00602(66) & 0.00157(66) & $1a$  \\[1mm]
	& Cu2 & $U_{\rm eqiv}$ (\r{A}$^2$) & 0.01025(75) & 0.00379(72) & $1b$  \\[1mm]
    & Cu3   & $z$ & 0.35816(16) & 0.35804(17)  &  $2q$  \\
	&		& $U_{\rm eqiv}$ (\r{A}$^2$) & 0.00531(45) & 0.00165(45)  \\[1mm]
	& Cu4    	& $z$ & 0.35609(16) & 0.35577(17) & $2s$  \\
	&		& $U_{\rm eqiv}$ (\r{A}$^2$) & 0.00554(45) & 0.00160(45) \\[1mm]
    & O1    & $z$ & 0.15882(77) & 0.15976(82)  &  $2q$  \\
	&		& $U_{\rm eqiv}$ (\r{A}$^2$) & 0.00871(238) & 0.00346(244)  \\[1mm]
    & O2    & $z$ & 0.15197(84) & 0.15279(85)  &  $2s$  \\
	&		& $U_{\rm eqiv}$ (\r{A}$^2$) & 0.01257(261) & 0.00564(249)  \\[1mm]
	& O3	& $x$ & 0.24981(54) & 0.25054(56) & $4w$  \\
	&		& $z$ & 0.37872(25) & 0.37855(32) &   \\
	&		& $U_{\rm eqiv}$ (\r{A}$^2$) & 0.00716(69) & 0.00300(87) \\[1mm]
    & O4    & $z$ & 0.37728(89) & 0.37697(99)  &  $2r$  \\
	&		& $U_{\rm eqiv}$ (\r{A}$^2$) & 0.00790(254) & 0.00740(290)  \\[1mm]
    & O5    & $z$ & 0.37955(87) & 0.37911(85)  &  $2t$  \\
	&		& $U_{\rm eqiv}$ (\r{A}$^2$) & 0.00650(244) & 0.00126(219)  \\[1mm]
	& O6    & $U_{\rm eqiv}$ (\r{A}$^2$) & 0.01463(258) & 0.00120(246) & $1e$  \\[1mm]
	&		& $R_1$ (\%) & 1.71 &  2.28 \\
	&		& $wR_2$ (\%) & 3.90 & 4.94 \\
	&		& GOF (\%) & 1.38 &  1.77 \\
			\end{tabular}
\end{ruledtabular}
\caption{\label{tab:table1} Crystallographic data for YBa$_{2}$Cu$_{3}$O$_{6.47(3)}$ at 295 and 100 K determined from single-crystal XRD.\@ The structure was refined in the orthorhombic space group $Pmmm$ using the doubled unit cell with the ortho-II chain oxygen ordering.\@  Due to space limitations only the equivalent atomic displacement parameters $U_{\rm eqiv}$ are listed, while the anisotropic atomic displacement parameters were used for the refinement. The corresponding Wyckoff positions (WP) are given as well.\@ }
\end{table}

\section{Momentum dependence of the elastic line}
\label{elast}
\begin{figure}
	\includegraphics[width=0.45\textwidth]{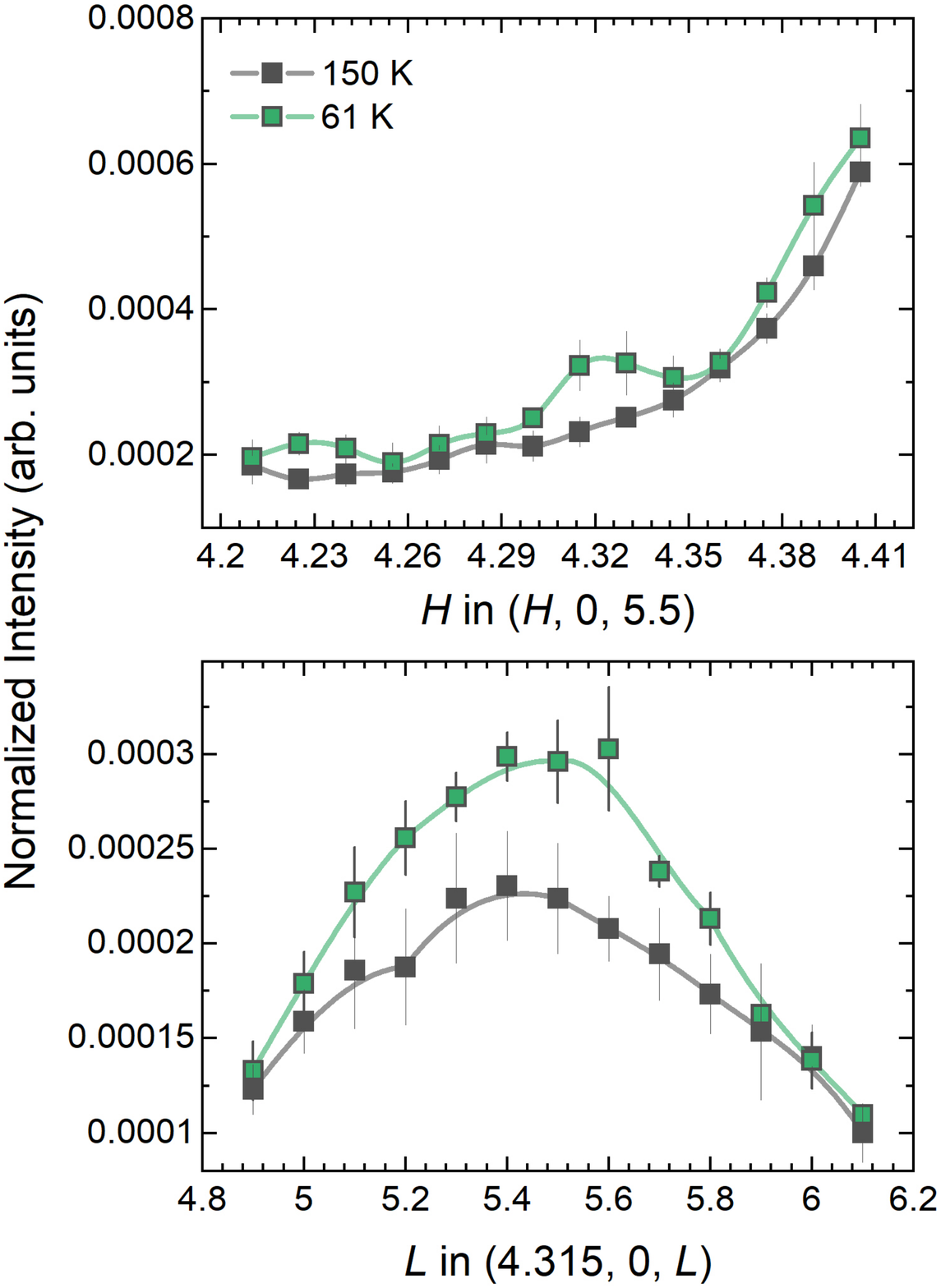}
	\centering
	\caption{(Color online) Momentum dependence of the quasi-elastic intensity measured at 150 K and at $T_c$ = 61 K, along the ($H$, 0, 5.5) and (4.315, 0, $L$) directions. The solid lines are guides to the eye.}
	\label{App_1}
\end{figure}

The momentum dependence of the elastic intensity across $Q_{2D}^a$ and along the $H$ and $L$ directions of the reciprocal space is plotted in Fig. \ref{App_1} and clearly shows that the intensity increase at $T_c$ is centered around $Q_{2D}^a$.
The CDW-related peak along the $H$ direction appears on top of a temperature-independent background of intensity increasing towards $H$ = 4.5, which is presumably related to the nearby ortho-II superlattice peak at Q$_{ortho-II}$ = (4.5 0 5).
It is worth mentioning that a broad peak centered at $Q_{2D}^a$ along the $L$ direction is already present in the data taken at 150 K, \textit{i. e.} at the CDW onset temperature. 
Further detailed temperature investigations would be required in order to identify the origin of this observation.
We note however that short-range dynamical charge density fluctuations have been reported in underdoped cuprates persisting up to temperatures well above the 2D-CDW onset ~\cite{Arpaia2019}.

\section{Density functional theory calculations}

Density-functional investigations were carried out for YBa$_2$Cu$_3$O$_7$ and YBa$_2$Cu$_3$O$_{6.5}$ using the mixed-basis pseudopotential method~\cite{Louie1979,Meyer}.
In both cases, the electron-ion interaction was represented by the same norm-conserving pseudopotentials, which included the semi-core states Y-4$s$, Y-4$p$, Ba-5$s$, Ba-5$p$, and O-2$s$ in the valence space.  In the mixed-basis approach valence states are expanded in a basis set consisting of a combination of plane waves and local functions. The latter allow an efficient description of more localized components of the valence states. Here, plane waves up to a kinetic energy of 20 Ry, augmented by local functions of $s$,$p$,$d$
type at the Y and Ba sites, of $s$ and $p$ type at the O sites and of $d$ type at the Cu sites were used. We employed the local-density approximation ~\cite{Hedin1971} for the exchange-correlation functional.

YBa$_2$Cu$_3$O$_{6.5}$ was modeled by the ordered ortho-II structure, where every second CuO chain is missing its oxygen.  The unit cell is thus doubled along the $a$ axis with respect to the  YBa$_2$Cu$_3$O$_7$ unit cell. Both orthorhombic structures were fully relaxed before calculating the phonon properties. Brillouin zone integrations were performed with a 12x12x4 and 4x6x2 orthorhombic k-point grid for YBa$_2$Cu$_3$O$_7$ and YBa$_2$Cu$_3$O$_{6.5}$, respectively,  in conjunction with a Gaussian smearing of 0.2 eV.  Further computational details can be found in  ~\cite{Bohnen2003,Heid2007}.

\begin{figure}
	\includegraphics[width=0.45\textwidth]{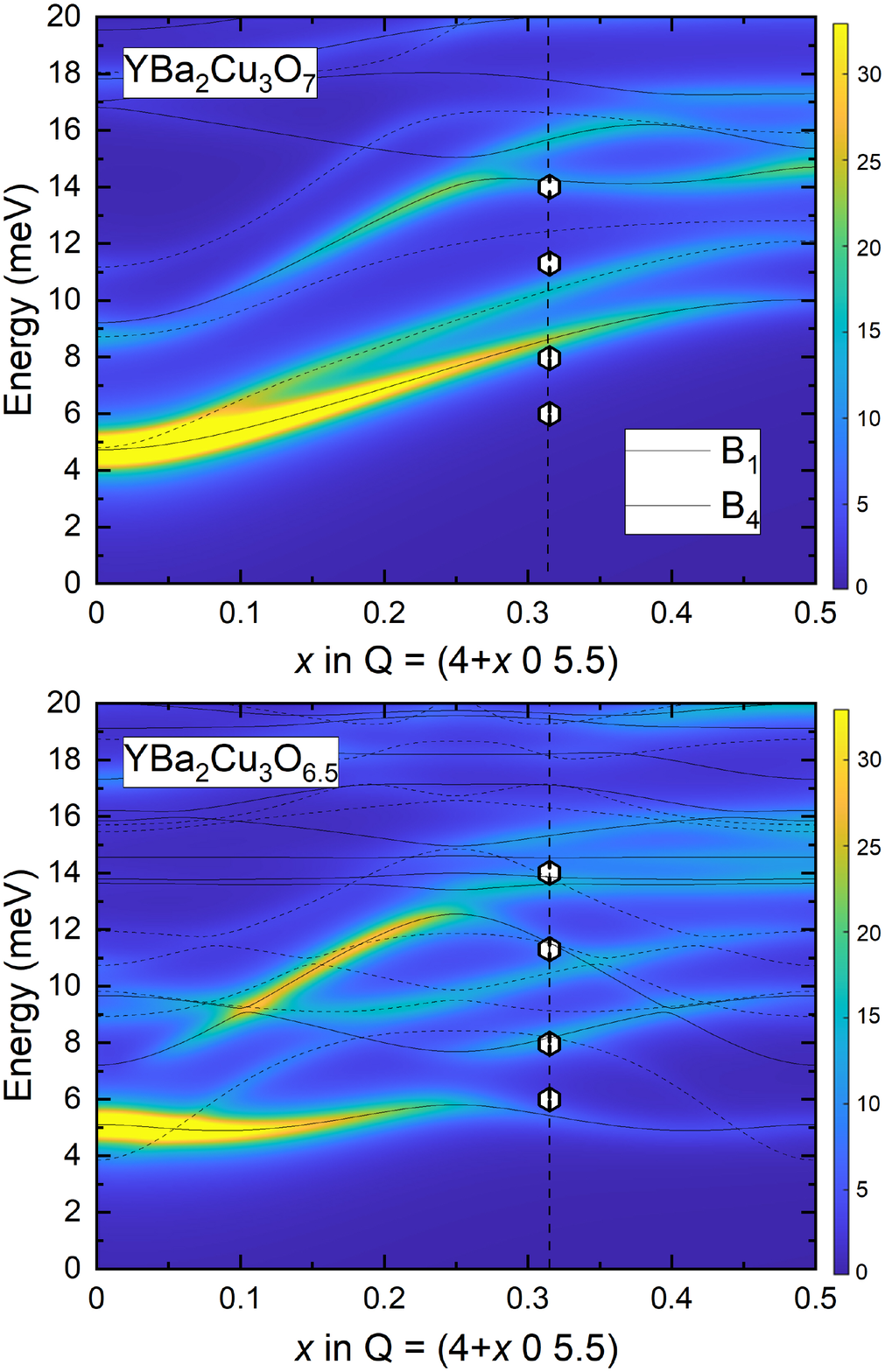}
	\centering
	\caption{(Color online) Colormap representation of the scattering intensity along the (x 0 0.5) direction and close to the G$_{405}$ = (4 0 5) Bragg peak calculated by density-functional perturbation theory for YBa$_2$Cu$_3$O$_7$ and YBa$_2$Cu$_3$O$_{6.5}$. The solid/dashed lines correspond to the calculated phonon dispersions of B$_{1}$/ B$_{4}$ symmetry branches respectively. The vertical dashed line indicates the $Q_{2D}^a$ position and the white symbols correspond to the results of the fitting analysis of IXS profiles, as described in the text. Note that for clarity, the reciprocal space positions are given in the notations of the ortho-I unit cell of YBa$_2$Cu$_3$O$_7$ for both compounds, even though the calculation for YBa$_2$Cu$_3$O$_{6.5}$ has been carried out with the unit cell doubled in the a direction.}
	\label{App_2}
\end{figure}

Lattice dynamics properties were calculated using the linear response or density-functional perturbation theory implemented in the mixed-basis scheme ~\cite{Heid1999}. Dynamical matrices were calculated on 4x4x2 on 2x2x2 orthorhombic grids for YBa$_2$Cu$_3$O$_7$ and YBa$_2$Cu$_3$O$_{6.5}$, respectively, and then obtained for arbitrary points in the Brillouin zone by standard Fourier-interpolation technique. Phonon dispersion and IXS structure factors were then calculated on the basis of the resulting phonon frequencies and eigenvectors. 

Along the reciprocal space direction (x 0 0.5), modes can be classified according to symmetry in four representations $B_i$ ($i = 1..4$). The number of modes for the two considered structures is listed in table~\ref{tab:modes} below. These representations further translate into symmetries of displacement patterns of the corresponding phonons listed in table ~\ref{tab:symmetry}.

\begin{table}[]
    \centering
    \begin{tabular}{|c|c|c|c|c|}
    \hline
    Structure  & $B_1$ & $B_2$ & $B_3$ & $B_4$\\
    \hline
    YBa$_2$Cu$_3$O$_7$   & 13 &6 &7 & 13\\
    \hline
    YBa$_2$Cu$_3$O$_{6.5}$ & 25 & 12 & 13 & 25\\
    \hline
    \end{tabular}
    \caption{Symmetry classification of the phonon modes along the (x, 0, 0.5) line for YBa$_2$Cu$_3$O$_7$ and YBa$_2$Cu$_3$O$_{6.5}$ crystal structures.}
    \label{tab:modes}
\end{table}

\begin{table}[]
    \centering
    \begin{tabular}{|c|c|c|c|c|}
    \hline
    Symmetry operation & $B_1$ & $B_2$ & $B_3$ & $B_4$\\
    \hline
    Identity   & + & + & + & +\\
    \hline
    $C_2(x)$ 180° rotation around x-axis  & + & + & - & -\\
    \hline
    $s(y)$   mirror symmetry y $\rightarrow$ -y & + & - & - & +\\
    \hline
    $s(z)$    mirror symmetry z $\rightarrow$ -z & + & - & + & -\\
    \hline
    \end{tabular}
    \caption{Symmetry operation for the displacement patterns in the different representations along the (x, 0, 0.5) line. 
    Note that the origin is taken at the Cu site of the CuO chains.}
    \label{tab:symmetry}
\end{table}
The calculated structure factors along the (x 0 0.5) direction and close to the G$_{405}$ = (4 0 5) Bragg peak are plotted in Fig. \ref{App_2} for YBa$_2$Cu$_3$O$_7$ and YBa$_2$Cu$_3$O$_{6.5}$. 
The agreement with the experimentally measured phonon peaks is better for the calculations done within the unit cell of the ortho-II ordered YBa$_2$Cu$_3$O$_{6.5}$.
For instance, the lowest energy phonon branch calculated for YBa$_{2}$Cu$_{3}$O$_{7}$ at $Q_{2D}^a$ (vertical line in Fig. \ref{App_2}) is at $\sim$9 meV.  
The experimentally observed phonon peak at $\sim$6 meV (Fig. \ref{f2}) is reproduced by the calculations only when the ortho-II doubling of the unit cell along the crystallographic a-axis and the resulting folding of the YBa$_{2}$Cu$_{3}$O$_{7}$ phonon dispersion curves is taken into account.

\section{Phonon anomalies along the $H$ direction}
\label{phonons_otherq}

\begin{figure}
	\includegraphics[width=0.42\textwidth]{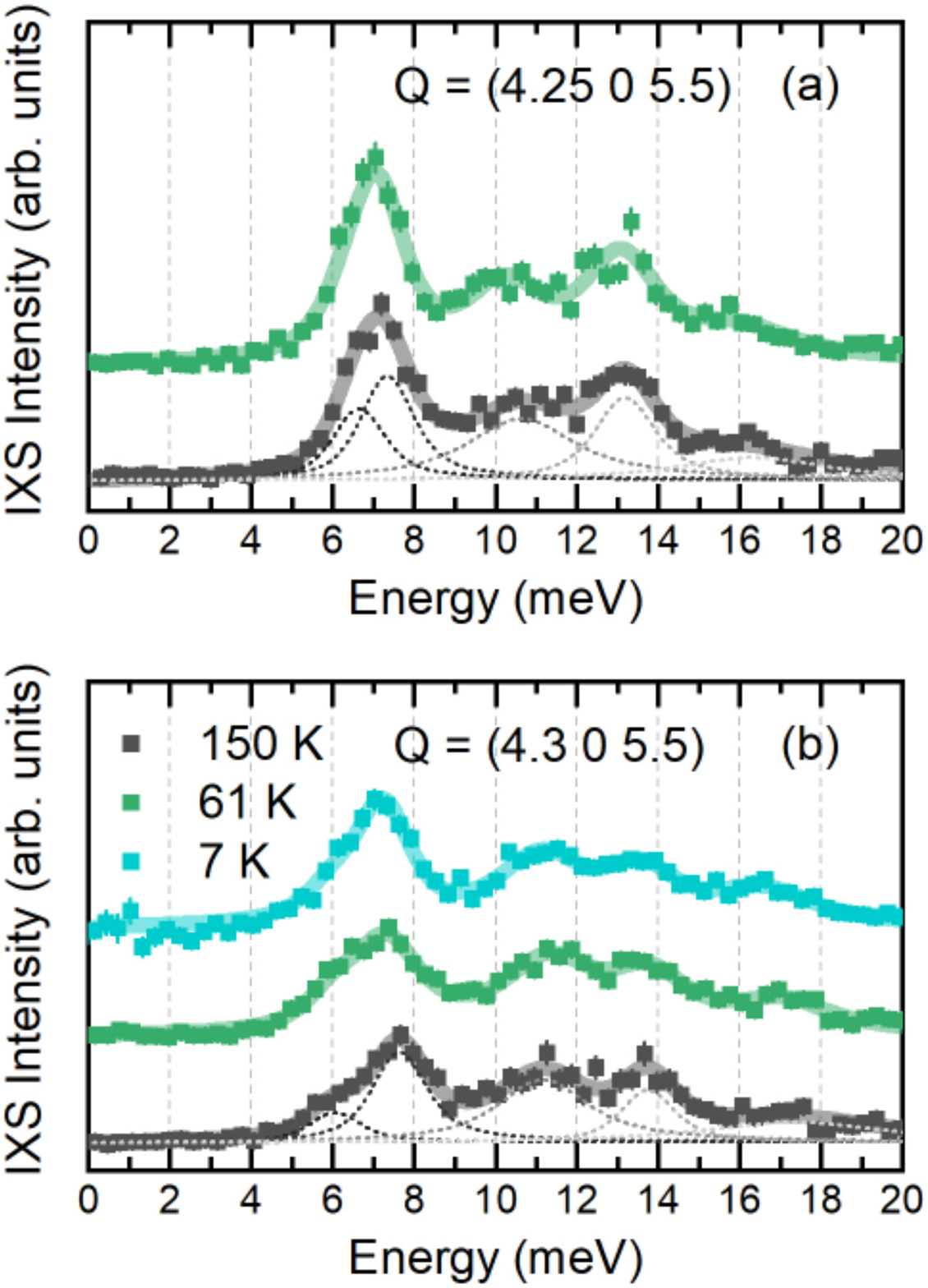}
	\centering
	\caption{(Color online) Inelastic part of the IXS spectra at Q = (4.25 0 5.5) and (4.3 0 5.5) at 150 K, at $T_c$ = 61 K and at 7 K. Thick solid lines correspond to the results of the least square fitting of the data and the dotted lines indicate the contributions from the individual fitted phonon profiles.}
	\label{App_3}
\end{figure}

IXS phonon spectra at Q = (4.25 0 5.5) and Q = (4.3 0 5.5), i.e. off from $Q_{2D}^a$ along the $H$ direction, are shown in Fig. \ref{App_3}. 
Within our resolution, we could not detect any marked anomalies at the spectra recorded at Q = (4.25 0 5.5) upon cooling from 150 K to $T_c$. 
At Q = (4.3 0 5.5) we observe a broadening and softening of the phonon at $\sim$8 meV, similar to the ones seen at $Q_{2D}^a$, albeit smaller in magnitude.

\section{IXS data at additional Brillouin zones}
\label{phonons_otherBZ}

\begin{figure}
	\includegraphics[width=0.45\textwidth]{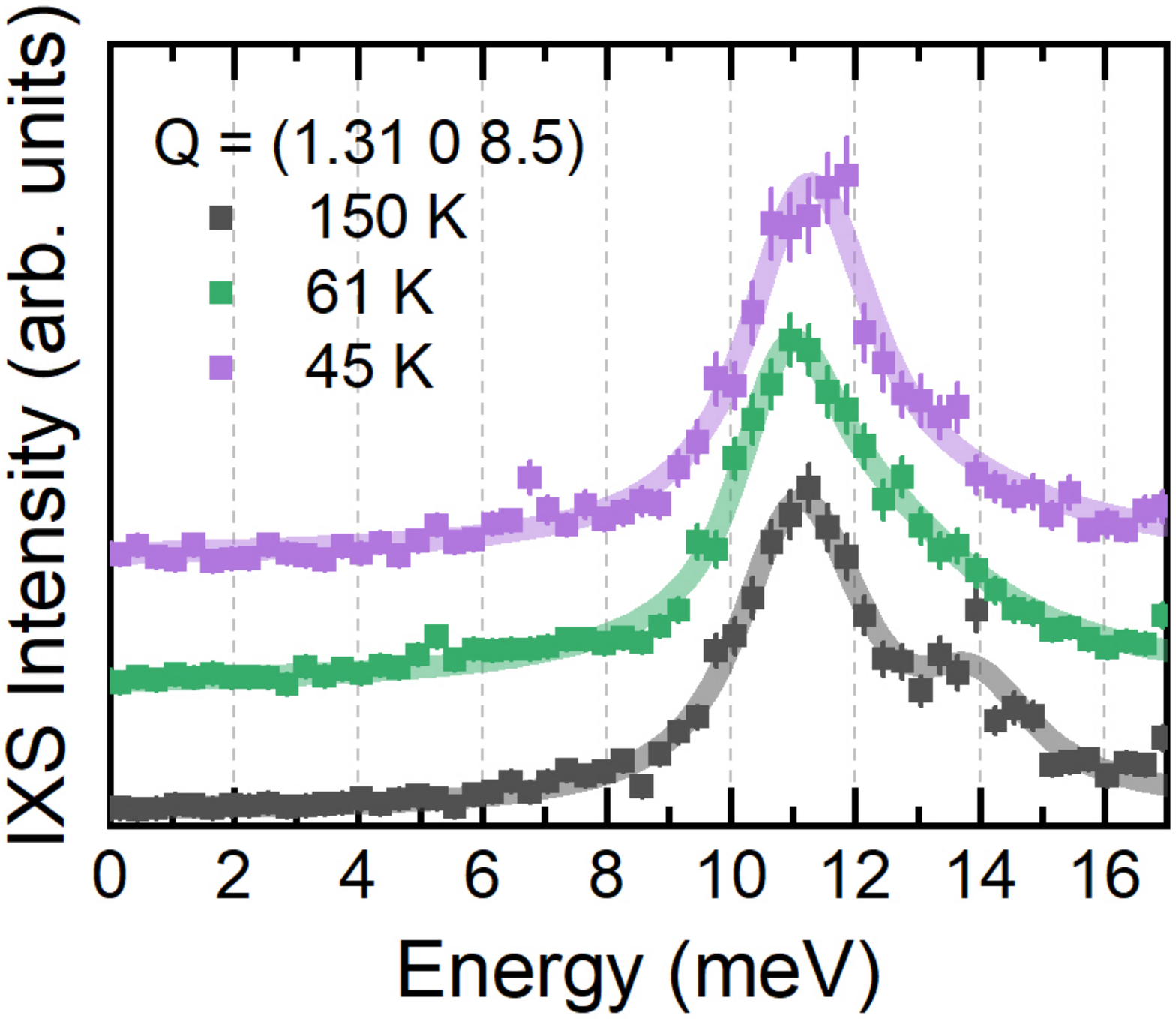}
	\centering
	\caption{(Color online) Inelastic part of the IXS spectra at Q = (1.31 0 8.5) at 150 K, at $T_c$ = 61 K and at 7 K. Thick solid lines correspond to the results of the least square fitting of the data }
	\label{App_4}
\end{figure}

In addition to the IXS measurements performed close to the G$_{405}$ = (4 0 5) Bragg reflection and presented in the main text, complementary measurements were performed close to the Brillouin zone adjacent to G$_{108}$ = (1 0 8) and are presented in Fig. \ref{App_4}.
The results of our density-functional perturbation theory calculations indicate that the strong phonon peak observed at $\sim$11 meV corresponds to a phonon branch belonging to the $B_{4}$ representation. 
In our IXS measurements we did not observe any significant changes on its linshape and energy upon cooling. 
Instead, the weaker phonon peak observed at $\sim$14 meV at 150 K - corresponding to a branch of the $B_{1}$ representation - changes upon cooling.
Although the intensity of this phonon peak is weak in this Brillouin zone, it is clear from the IXS data that the phonon softens at low temperatures. 

\end{document}